\documentclass{JHEP3}
\usepackage{amsmath,amssymb}
\usepackage{graphicx,epsfig}
\usepackage{wasysym}

\newcommand{\fs}{\; .}
\newcommand{\co}{\; ,}  
\newcommand{\non}{\nonumber}
\newcommand{\no}{\nonumber \\}
\newcommand{\quadp}{\hspace{1.3em}}
\newcommand{\eps}{0}
\newcommand{\M}{M}
\newcommand{\mh}{\hat{m}}
\newcommand{\Fz}{F_0}
\newcommand{\mpi}{m_\pi}
\newcommand{\mK}{m_K}
\newcommand{\meta}{m_{\eta}}
\newcommand{\mP}{m_P}
\newcommand{\Lpi}{l^r_\pi}
\newcommand{\LK}{l^r_K}
\newcommand{\Leta}{l^r_\eta}
\newcommand{\LP}{l^r_P}
\newcommand{\Fpi}{F_\pi}
\newcommand{\Mpi}{M_\pi}
\newcommand{\MK}{M_K}
\newcommand{\Meta}{M_{\eta}}
\newcommand{\MP}{M_P}
\newcommand{\LLpi}{L^r_\pi}

\newcommand{\LLeta}{L^r_\eta}
\newcommand{\LLP}{L^r_P}
\newcommand{\cloop}{\bar{c}_{\rm loop}}
\newcommand{\cloglog}{c_{\log \times \log}}
\newcommand{\clog}{c_{\log}}
\newcommand{\clogLi}{c_{\log \times  L_i}}
\newcommand{\cLi}{c_{L_i}}
\newcommand{\cLiLj}{c_{L_i \times L_j}}
\newcommand{\cCi}{c_{C_i}}
\preprint{CPT-P32-2007\\
July 2007 (v1)\\
September 2007 (v2)}

\title{On the two-loop contributions to the pion mass}

\author{Roland Kaiser\\
Centre de Physique Th\'{e}orique,\footnote{Unit\'{e} mixte de
recherche (UMR 6207) du CNRS et des Universit\'{e}s Aix-Marseille I,
Aix-Marseille II, et du Sud Toulon-Var; laboratoire affili\'{e} \`{a}
la FRUMAM (FR 2291).}  CNRS-Luminy,\\ 
Case 907, F-13288 Marseille Cedex 9, FRANCE\\
E-mail:  \email{kaiser@cpt.univ-mrs.fr}
}

\abstract{
We derive a simplified representation for the pion mass to two loops in three-flavour chiral perturbation theory. For this purpose, we first determine the reduced expressions for the tensorial two-loop 2-point sunset integrals arising in chiral perturbation theory calculations. Making use of those relations, we obtain the expression for the pion mass in terms of the minimal set of master integrals. On the basis of known results for these, we arrive at an explicit analytic representation, up to the contribution from $ K K \eta $ intermediate states where a  closed-form expression for the corresponding sunset integral is missing. However, the expansion of this function for a small pion mass leads to a simple representation which yields a very accurate approximation of this contribution. Finally, we also give a discussion of the numerical implications of our results. 
}

\keywords{Chiral Lagrangians, NLO Computations, Spontaneous Symmetry Breaking, Nonperturbative Effects.
\\[0.3em]11.15.Bt, 11.30.Rd, 12.38.Bx}

\begin{document}

\section{Introduction}

Calculations in chiral perturbation theory~\cite{Weinberg:1978kz,Gasser:1983yg,Gasser:1984gg} beyond one loop have proven to be intricate. One of the difficulties lies in the contributions from genuine two-loop integrals involving massive internal lines. In the case of three-flavour chiral perturbation theory these can in addition involve more than one mass scale which so far often prohibited the analytic evaluation. 
For example, an analytical expression for the pion mass in this framework is not available. 
In the present paper, we derive an explicit representation for the two-loop correction to the pion mass in terms of dilogarithms. 
Our final result is still not entirely analytic because the closed-form expression for the $KK\eta$ intermediate states is missing. We account for these contributions in terms of a simple approximate representation suitable for all practical purposes including investigations of the quark mass dependence.  

In the two-flavour version of chiral perturbation theory \cite{Gasser:1983yg} the pion mass to two loop order was calculated in Ref.~\cite{Burgi:1996qi} and further useful representations of the result can be found in Refs.~\cite{Bijnens:1997vq} and \cite{Bijnens:1998fm}. In this case, the calculation only involves equal mass on-shell sunset integrals which were evaluated analytically.
The two-loop calculation of the pion mass in three-flavour chiral perturbation theory~\cite{Gasser:1984gg} was performed in Refs.~\cite{Golowich:1997zs,Amoros:1999dp}.
These calculations were performed in the isospin limit $m_u =m_d$. A calculation of isospin breaking effects in the meson masses is described in Ref.~\cite{Amoros:2001cp}.   
For a recent review of chiral perturbation theory beyond one loop we refer to Ref.~\cite{Bijnens:2006zp} and an extensive list of references can also be found in Ref.~\cite{Kaiser:2006uv}. A detailed description of the calculational methods used in the framework of the two-flavour theory is provided in Ref.~\cite{Gasser:1998qt}. Recently, explicit analytic results involving the strange quark flavour have been obtained for the two-loop matching of the low-energy constants $ B$~\cite{Kaiser:2006uv}, $ F$ and $ \ell_1 , \ldots , \ell_7 $~\cite{Gasser:2007sg}. The analogous relation for the strange quark mass dependence of the up-quark condensate $\langle 0 | \bar{u}u | 0 \rangle $ was derived in Ref.~\cite{Moussallam:2000zf}. In this last case, there are no contributions from sunset integrals.   

On the other hand, the final results for the pion mass in Refs.~\cite{Golowich:1997zs,Amoros:1999dp} involve contributions from sunset integrals that have not been evaluated analytically. Among these, there are furthermore tensorial sunset integrals of rank 1 and 2.\footnote{In the two-flavour case, the tensorial integrals were evaluated along with the scalar integral~\protect\cite{Gasser:1998qt}.} 
By Lorentz invariance, such tensorial integrals may always be related to linear combinations of scalar integrals \cite{Passarino:1978jh}. In the case of a one-loop integral, the scalar products involving the loop momentum in the numerator may then always be re-expressed in terms of the denominators, masses and scalar products of the external momenta. Thus, the procedure leads to a representation of any tensorial integral in terms of the corresponding scalar integral plus simpler functions. In the case of the two-loop 2-point sunset integral this procedure does not go through, because the number of scalar products that can be formed with the  two loop momenta and the external one (five) is greater than the number of available denominators (three). In general, one thus ends up with additional integrals with irreducible numerators. 

In Refs.~\cite{Tarasov:1996br,Tarasov:1997kx} Tarasov has pointed out that such integrals with irreducible numerators can always be related to integrals with unit numerators, at the price of shifting the dimension $d$ in dimensional regularization by multiples of 2. This then calls for finding a connection of integrals with unit numerators in different dimensions. In fact, such relations can be found easily from the Feynman parameter representation for a given integral. On the basis of these observations, Tarasov shows that, up to contributions from tadpoles, any (tensorial) sunset integral with denominators raised by arbitrary positive integer powers can recursively be expressed as a linear combination of four master integrals, with coefficients formed with the three masses, the momentum and the dimension $d$. The four master integrals consist of the sunset integral with unit powers of the denominators plus those three where in each case one of the denominators appears squared. The recurrence relations of Ref.~\cite{Tarasov:1997kx} have been tested and implemented in a computer program package in Ref.~\cite{Mertig:1998vk}. For a review of the `integration by parts' relations \cite{'tHooft:1978xw,Chetyrkin:1981qh,Vassiliev} that are at the heart of these results we refer the reader to Ref.~\cite{Smirnov:2004ym} which also  contains further bibliographical references. The problem of (massive) tensorial two-loop integrals is also considered in Refs.~\cite{Weiglein:1993hd,Ghinculov:1994sd,Post:1996gg,Ghinculov:1997pd,Harlander:1998dq,Ghinculov:2000cz,Actis:2004bp}.       

As an application of the general results of Ref.~\cite{Tarasov:1997kx}, we derive the reduced expressions for the tensorial sunset integrals relevant for chiral perturbation theory calculations. Their representation in terms of the four master integrals is provided for the general case with arbitrary momentum and three different masses. With these expressions at hand, we then work out a simplified representation for the pion mass to two loops in three-flavour chiral perturbation theory. Our derivation relies on the representation given in Ref.~\cite{Amoros:1999dp} where the contributions of the coupling constants of the order $p^6$ chiral Lagrangian are given in the standard basis determined in Ref.~\cite{Bijnens:1999sh}, for related work see also Ref.~\cite{Haefeli}. 
Our explicit result shows that the consistent reduction to master integrals leads to a final result in terms of fewer basis functions compared to Ref.~\cite{Amoros:1999dp}. 

An account of analytically known special cases of two-loop 2-point functions can be found in Refs.~\cite{Davydychev:1999ic,Martin:2003qz,Martin:2005qm}. For recent analytic results for the equal mass sunset graph we refer to Refs.~\cite{Laporta:2004rb,Tarasov:2006nk}. 
For the pion mass we arrive at an explicit representation in terms of dilogarithms, involving in the case of the contribution from the $KK\eta$ intermediate states an expansion for a small pion mass. Our result 
is presented in such a manner that the renormalization scale dependence of the various contributions is manifest: For example we separately display the contributions involving double and single chiral logarithms and one separate term accounts for the finite and scale invariant contributions from the sunset integrals. The advantage of such a representation is exploited in the subsequent numerical study of our results, where we in particular investigate the scale variations of the individual contributions. Previous numerical investigations of the two-loop corrections to the pion mass can be found in Refs.~\cite{Amoros:1999dp,Amoros:2001cp,Amoros:1999qq,Amoros:2000mc,Bijnens:2002hp,Bijnens:2003uy,Bijnens:2003xg,Bijnens:2004eu,Bijnens:2004bu} and a brief overview of these is given in Ref.~\cite{Bijnens:2007si}.

The paper is organized as follows: The recurrence relations and results for the tensorial sunset integrals are presented in Section~\ref{tensorial sunsets}. The dimension $d$ is left arbitrary throughout this section. In Section~\ref{Renormalization and results}, we then provide the relations required to separate the divergent and finite parts in the vicinity of $d =4$ and give explicit expressions for the finite parts for a number of special cases of interest. 
In Section~\ref{pion mass}, these results are applied to yield a simplified representation of the pion mass. The algebraic evaluation is followed by a discussion of the numerical implications of our results in Section~\ref{numerics}.  Finally, Section~\ref{Conclusions} contains our conclusions. For the purpose of illustration, we provide in Appendix~\ref{App:recrel-one-loop} a discussion of the recurrence procedure in the case of the one-loop 2-point function. A second Appendix~\ref{App:Sunset for s=0} demonstrates the absence of spurious singularities in the limit of vanishing external momentum in the reduced expressions for the tensorial sunset integrals.

\section{Recurrence relations for the sunset integral}
\label{tensorial sunsets}

\begin{figure}[t]
\begin{center}
\epsfig{figure=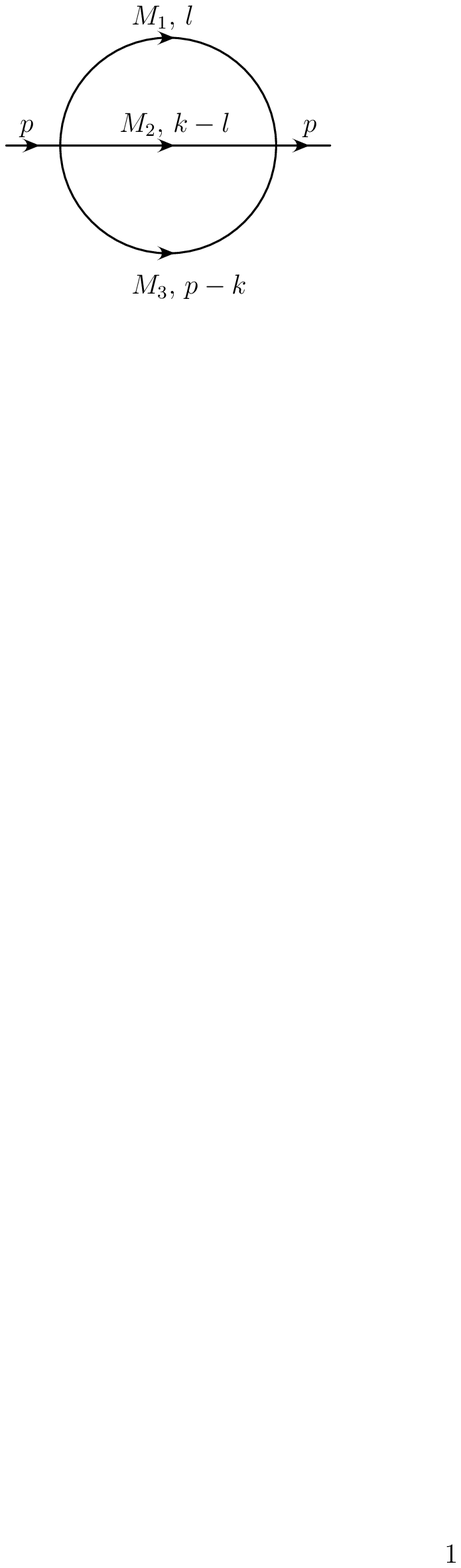,height=4.3cm,bbllx=90,bblly=630,bburx=240,bbury=765,clip}
\end{center}
{\caption{The sunset topology.}
\label{fig_sunset}}
\end{figure}

We consider the family of two-loop integrals defined by
\begin{align}
& S^{d}_{\alpha,\beta,\gamma}  \{ \M_1 , \M_2 ,\M_3 ; p^2  \} =  
\\
 & \quad \frac{1}{i^2} \!  \int \! \frac{d^d k}{(2\pi)^d}\frac{d^dl}{(2\pi)^d} \frac{1}{[M_1^2-l^2-i \eps]^\alpha} \frac{1}{[M_2^2-(k-l)^2-i \eps]^\beta} \frac{1}{[M_3^2-(p-k)^2-i \eps]^\gamma} 
\fs  \non
\end{align}
These integrals are often referred to as `sunset' integrals, in allusion to the corresponding Feynman graph shown in Figure~\ref{fig_sunset}. Making use of standard techniques, we arrive at the following Feynman parameter integral representation for the sunset (see e.g. Ref~\cite{Kaiser:2006uv})
\begin{align}
\label{FeynPSunset}
 S^d_{\alpha,\beta,\gamma} \{M_1, M_2 , M_3  ; s \} =   \frac{1}{(4 \pi)^{d}} \frac{\Gamma(\alpha+\beta+\gamma-d)}{\Gamma(\alpha)\Gamma(\beta)\Gamma(\gamma)} \int_0^1 \! dx \, x^{\alpha+ \gamma - \frac{d}{2} -1} \bar{x}^{\beta + \gamma- \frac{d}{2}-1}     
\\
 \times     \int_0^1 \! dy \, y^{\alpha+\beta- \frac{d}{2}-1} \bar{y}^{\gamma-1}  [ x y  M_1^2 + \bar{x} y M_2^2 + x \bar{x} \bar{y} (M_3^2  -y s ) - i \eps ]^{d-\alpha-\beta-\gamma} 
\co \non
\end{align}
where $\bar{x} = 1- x$, etc. The form of this representation suggests the existence of relations among integrals with different indices $ \alpha , \beta $ and $ \gamma $. In fact, Tarasov \cite{Tarasov:1997kx} has shown that all the sunset integrals with positive integer indices may be expressed in terms of the 4 integrals satisfying $ \alpha +\beta + \gamma \leq 4 $ plus products of tadpole integrals,  
\begin{align}
I^d_\alpha \{ M \} =  \frac{1}{i} \!  \int \! \frac{d^d k}{(2\pi)^d}  \frac{1}{[M^2-k^2-i \eps]^\alpha}  = \frac{1}{(4 \pi)^\frac{d}{2}} \frac{\Gamma(\alpha-\frac{d}{2})}{\Gamma(\alpha)} \, M^{d -2 \alpha} 
\fs
\end{align} 
The recurrence relations for the sunset integrals are quite involved and shall not be repeated here. For the purpose of illustration we list the analogous relations for the one-loop 2-point function in Appendix~\ref{App:recrel-one-loop}. For the tadpole, the recurrence relation reads,
\begin{align}
I^d_\alpha \{ M \}  =  \frac{\alpha-\frac{d}{2}-1}{(\alpha-1)M^2 } \, I^d_{\alpha-1} \{ M \} \co   \quad   \alpha \neq 1 \fs  
\end{align}

For specific configurations of masses and momentum there are 
additional relations leading to a reduced number of master integrals, 
\begin{align}
\label{SmmMM}
S^d_{1,1,2}\{\M_1,\M_1,\M_2 ; \M_2^2\} & = - \frac{3d-8}{4 \M_2^2} \,S^d_{1,1,1}\{\cdot \}   - \frac{\M_1^2}{\M_2^2} \,S^d_{2,1,1}\{ \cdot \} - \frac{(d-2)^2 }{8(d-3)}  \frac{ I^d_1\{ \M_1 \}^2 }{\M_1^2 \M_2^2 }
\co \\
\label{Smmmm}
 S^d_{2,1,1}\{M,M,M ; M^2\} & = - \frac{3d-8}{8 M^2} \,S^d_{1,1,1}\{\cdot \} - \frac{(d-2)^2}{16(d-3)M^4}\, I^d_1\{M \}^2 
\fs 
\end{align}
Here and in the following, the symbol $\{\cdot \}$ appearing on the r.h.s. of an equation is a shortcut for the repetition of the set of arguments introduced on its l.h.s. The relation Eq.~(\ref{SmmMM}) 
was given in Ref.~\cite{Onishchenko:2002ri} and Eq.~(\ref{Smmmm}) was discussed in Ref.~\cite{Mertig:1998vk}, but both may also be inferred from the general expressions given in Ref.~\cite{Tarasov:1997kx}.  

In general, contributions of graphs with indices $ \alpha, \beta , \gamma $ other than unity 
are not typical and chiral perturbation theory is no exception.
However, due to the derivative couplings in the chiral Lagrangian one commonly encounters 
tensorial integrals.\footnote{See for example Refs.~\cite{Golowich:1997zs,Amoros:1999dp}.} 
Below, we will show how these may be reduced to the same set of master integrals. 
It is convenient to consider the tensorial integrals associated with the totally symmetric traceless tensors  $ \tau^{\mu_1 \cdots \mu_n}(l) $,
\begin{align} 
& \tau^{\mu}(l) = l^\mu   
\co \quad 
\tau^{\mu\nu}(l) = l^\mu  l^\nu - \frac{l^2}{d}  \, g^{\mu \nu}    
\co \\
& \tau^{\mu\nu\rho }(l) = l^\mu  l^\nu l^\rho - \frac{l^2}{d+2}  ( g^{\mu \nu} l^\rho +g^{ \nu \rho } l^\mu   +g^{\rho \mu } l^\nu )
\co \quad {\rm etc.}
 \non
\end{align}
For $n =1,2,3 , \ldots  $ we then define the functions $S^{(n)d}_{\alpha,\beta,\gamma}  \{ \M_1 , \M_2 ,\M_3 ; p^2  \} $,\footnote{Note that in our notation, the superscript `$(n)$' does not indicate a momentum derivative.}   
\begin{align}
\label{tensorintegrals}
& \tau^{\mu_1 \cdots  \mu_n}(p)\,   S^{(n)d}_{\alpha,\beta,\gamma}  \{ \M_1 , \M_2 ,\M_3 ; p^2  \}   = 
\\
 & \quad \quad  \frac{1}{i^2} \!  \int \! \frac{d^d k}{(2\pi)^d}\frac{d^dl}{(2\pi)^d} \frac{\tau^{\mu_1 \cdots  \mu_n}(l)}{[M_1^2-l^2-i \eps]^\alpha} \frac{1}{[M_2^2-(k-l)^2-i \eps]^\beta} \frac{1}{[M_3^2-(p-k)^2-i \eps]^\gamma} 
 \fs \non
\end{align}  
The validity of Eq.~(\ref{tensorintegrals}) can be verified, e.g., via the same steps that lead to the Feynman parameter integral representation in Eq.~(\ref{FeynPSunset}) and by using the relation
\begin{align}
\label{therelation}
\int \! d^d l  \, \tau^{\mu_1 \cdots  \mu_n}(l + x  p )\, f (l^2) = \tau^{\mu_1 \cdots  \mu_n}( p ) \, x^n \! \int \! d^d l\,  f (l^2) 
\fs
\end{align}
In doing so, we moreover arrive at a general result for the tensorial integral of rank $n$,
\begin{align}
\label{tensor_n}
S^{(n)d}_{\alpha,\beta,\gamma}  \{ \M_1 , \M_2 ,\M_3 ; p^2  \}  = (4\pi)^{2n}\,  \frac{\Gamma(\beta+n)\Gamma(\gamma+n)}{\Gamma(\beta)\Gamma(\gamma)} \, S^{d+2n}_{\alpha,\beta+n,\gamma+n}  \{ \cdot   \}  \fs
\end{align}
This relation exemplifies a general finding of Tarasov \cite{Tarasov:1996br,Tarasov:1997kx}: Tensorial integrals can be expressed in terms of scalar integrals with the dimension $d$ shifted by multiples of $2$. It remains to find a relationship between scalar integrals in shifted dimensions. For the tadpoles, these relations are again very simple,
\begin{align}
I^d_1 \{ M \}  = -  \frac{M^2 }{2\pi (d-2)} \, I^{d-2}_{1} \{ M \} \co   \quad   d  \neq 2 \fs  
\end{align}
In the case of the sunset, a suitable relation can be obtained from the Feynman parameter integral representation in Eq.~(\ref{FeynPSunset}). By inserting the identity $1 = y  +(1-x)(1-y)  + x (1-y)$ under the integral signs, we deduce~\cite{Tarasov:1997kx}
\begin{align}
& S^{d}_{\alpha,\beta,\gamma} \{ \M_1 , \M_2 ,\M_3 ; s  \}  = 
\\
&\quad \quad\quad 
(4\pi)^2 \left[ \alpha \, \beta\,  S^{d+2}_{\alpha+1 ,\beta+1 ,\gamma} \{  \cdot \} + \beta\,  \gamma\,  S^{d+2}_{\alpha  ,\beta+1 ,\gamma+1 } \{  \cdot \}  +\gamma \, \alpha\,  S^{d+2}_{\alpha+1 ,\beta ,\gamma+1 } \{  \cdot \}  \right] \non 
\fs
\end{align}
On the basis of this equation evaluated for $ \alpha + \beta + \gamma \le 4 $ and subsequent index reduction one can derive the linear relations between the master integrals in $d $ and in $d-2 $ dimensions. Again, the formal expressions given explicitly in Ref.~\cite{Tarasov:1997kx} are quite involved. Making use of these results and Eq.~(\ref{tensor_n}) we arrive at the comparatively simple relations for the tensorial integrals with unit indices, 
\begin{align}
\label{S1}
& S^{(1)d}_{1,1,1} \{ \M_1 , \M_2 ,\M_3 ; s  \} = 
\\
& \quad \frac{(d-2)s - (d-3) (2 \M_1^2- \M_2^2 - \M_3^2)}{3(d-2)s} \, S^{d}_{1,1,1} \{\cdot\}  +  2 \M_1^2 \frac{s - \M_1^2}{ 3 (d-2)s}  \, S^{d}_{2,1,1} \{\cdot \} 
\no
& \quad -  \M_2^2 \, \frac{s - \M_2^2+3 (\M_1^2-\M_3^2)}{ 3 (d-2)s}  \, S^{d}_{1,2,1} \{\cdot  \}   - \M_3^2 \, \frac{s - \M_3^2+3 (\M_1^2-\M_2^2)}{ 3 (d-2)s}  \, S^{d}_{1,1,2} \{ \cdot  \} 
\no 
& \quad+\frac{1}{6s} \, I^d_1 \{ \M_1\} [ I^d_1 \{ \M_2\}+  I^d_1 \{ \M_3 \}   ] - \frac{1}{3 s} \, I^d_1 \{ \M_2\}  I^d_1 \{ \M_3 \} 
\co\non\end{align} \begin{align}
\label{S2}
& S^{(2)d}_{1,1,1} \{ \M_1 , \M_2 ,\M_3 ; s  \} = 
\\
&  \quad \frac{D}{ds^2} \,  \Big\{  2d (d-1)(d-2)s^2 +(d-3) [4 (d-1) (d-4) \M_1^2 +d(7d-12)\Sigma_{2,3} ] s 
\non \\
& \hspace{2.5em}  - d(d-3) [ 4 (d-1) \M_1^4 - (d+4) \M_1^2  \Sigma_{2,3} + d \Sigma_{2,3}^2 +12 (d-2) \M_2^2 \M_3^2 ]   \Big\}  \, S^{d}_{1,1,1} \{ \cdot  \}  
\non \\ 
&    \quad + 4 D \M_1^2  \Delta_{s,1}  \, \frac{ (d-1) \Sigma_{s,1} + (d-3)  \Sigma_{2,3} }{s^2} \, S^{d}_{2,1,1} \{ \cdot  \}  
\non \\ 
&   \quad - \frac{D \M_2^2}{s^2 } \,  \Big\{  2 [2(d-2)s+(d+2)\M_1^2 -(11d-18) \M_3^2] \Delta_{s,2} 
\non \\
& \hspace{13.5em}  + d \Delta_{s,2}^2 +3(3d-4) \Delta_{1,3}^2   \Big\}  \, S^{d}_{1,2,1} \{ \cdot  \}  + (\M_2  \leftrightarrow \M_3) 
\non \\ 
&   \quad + D(d-2) \, \frac{ (5d-8) \Sigma_{s,1} -d \M_2^2 - (7d-12) \M_3^2  }{2s^2}  \, 
I^d_1 \{ \M_1\}  I^d_1 \{ \M_2\} + (\M_2  \leftrightarrow \M_3)
\non \\ 
&  \quad -  D(d-2) \, \frac{ (11 d^2 -32 d +24 ) s  + d (5d-8) \M_1^2 - 2d (2 d-3) \Sigma_{2,3}  }{d s^2}  \, 
I^d_1 \{ \M_2 \}  I^d_1 \{ \M_3 \} 
\fs \non
\end{align}
We used the following abbreviations
\begin{alignat}{3}
D & = \frac{d}{6(3d-4)(d-1)(d-2)}  \co \quad &\Sigma_{i,j} & = \M_i^2 + \M_j^2  \co \quad &  \Delta_{i,j}  &= \M_i^2 - \M_j^2  \co 
\\
&& \Sigma_{s,j} & = s + \M_j^2 \co \quad   &  \Delta_{s,j} & = s-  \M_j^2  
\fs \non   
\end{alignat}
We have checked Eqs.~(\ref{S1}) and (\ref{S2}) using the computer program package of Ref.~\cite{Mertig:1998vk}.
Note that apart from the increasing length of the resulting expressions, there arise no complications in the evaluation of the tensorial integrals of rank $n$ greater than 2. These are, however, not required for the envisaged application.    
In the case of equal masses, $ \M_1 = \M_2 = \M_3  = M $, the relation in Eq.~(\ref{S1}) simplifies to  
\begin{align}
 S^{(1)d}_{1,1,1}\{M,M,M ; s\} & = \frac{1}{3} \,S^d_{1,1,1}\{\cdot \}   
\fs
\end{align}
This relation was already employed in Ref.~\cite{Amoros:1999dp} in order to achieve a simplification in their expressions. 

\section{Renormalization and explicit results for the master integrals}
\label{Renormalization and results}

The divergent part of the sunset integral is known for arbitrary values of masses and momenta~\cite{vanderBij:1983bw,Ramond,Caffo:1998du}. For $s \ge 0$, it is convenient to define a quantity $S^{{\rm div}}_{1,1,1} \{M_1 ,M_2, M_3; s \}$,
\begin{align}
\label{Eq: Sdiv def}
& S^{{\rm div}}_{1,1,1} \{M_1 ,M_2, M_3; s \} = \frac{\mu_\chi^{2d-8}}{(4\pi)^4} \bigg\{ -\frac{2}{(d-4)^2} \, (\M_1^2 + \M_2^2 + \M_3^2 )
\\
&  \quad - \frac{1}{(d-4)} \left[  
\M_1^2  \left(  2 \ln \tfrac{\M_1^2}{\mu^2} -1\right) + 
\M_2^2  \left(  2 \ln \tfrac{\M_2^2}{\mu^2} -1\right) + 
\M_3^2  \left(  2 \ln \tfrac{\M_3^2}{\mu^2} -1\right)  + \frac{s}{2} \right] 
\no
& \quad 
- \M_1^2   \ln \tfrac{\M_1^2}{\mu^2} \left(  \ln \tfrac{\M_1^2}{\mu^2} -1\right) 
-  \M_2^2   \ln \tfrac{\M_2^2}{\mu^2} \left(  \ln \tfrac{\M_2^2}{\mu^2} -1\right) 
-   \M_3^2   \ln \tfrac{\M_3^2}{\mu^2} \left(  \ln \tfrac{\M_3^2}{\mu^2} -1\right) 
- \frac{s}{2} \ln \frac{s}{ \mu^2} \bigg\} 
\fs \non
\end{align}
As usual, the separation of the divergent parts involves an arbitrary scale $\mu $, which also enters in the combination $ \mu_\chi $ customary in chiral perturbation theory \cite{Gasser:1984gg},
\begin{align}
 \mu_\chi^2 = \frac{e^{-\Gamma'(1)-1}}{4\pi}\, \mu^2
\fs 
\end{align}
Our definition of the divergent piece in Eq.~(\ref{Eq: Sdiv def}) is such that the $\mu$-dependence of the contributions that explode in the limit $d \to 4$ is cancelled by a finite term, given in the third line, up to contributions of higher order in $ d-4$, viz. $ d S^{\rm div}_{1,1,1} \{ \cdot \}/d\mu  = O(d-4)$. The definition
\begin{align}
\label{Eq: Sbar def}
\bar{S}_{1,1,1} \{M_1 ,M_2, M_3; s \}  = \lim_{d \to 4} \left[  S^d_{1,1,1} \{ \cdot \}  -  S^{\rm div}_{1,1,1} \{ \cdot \}  \right] 
\fs  
\end{align}
thus implies that the resulting expression $ \bar{S}_{1,1,1} \{M_1 ,M_2, M_3; s \} $ is both finite and independent of the renormalization scale $ \mu $. 
The corresponding finite parts for the remaining master integrals are obtained from the above relation by differentiation with respect to the masses $M_1$, $M_2$ or $M_3$. 

We now turn to a number of special cases for which the analytic representation is also known for the finite parts and will be useful in the application below. For two equal masses and the momentum on the mass shell of the third mass the result is \cite{Berends:1997vk,Argeri:2002wz}, 
\begin{align}
\label{Eq:Sunset MMm on-shell}
& \bar{S}_{1,1,1} \{M ,M, m ; m^2 \} =
\\
& \quad  \frac{\M^2} {(4\pi)^4}   \left\{ \frac{(1-\tau)^2}{\tau}  \left[ {\rm Li}_2 (1 - \tau ) - \frac{\pi^2}{6} \right]  +\frac{\tau}{2}  \left[ \ln^2  \tau  + \frac{9}{4} \right]    -  \ln \tau  - (2 + \tau)  \left[   \frac{\pi^2}{12} + 1 \right]   \right\} 
\co \non 
\end{align}
where ${\rm Li}_2(z)$ denotes the dilogarithm, ${\rm Li}_2(z)=  \sum_{n=1}^\infty \frac{z^n}{n^2} \; (|z| \leq 1)$~\cite{Lewin}.
Note that the above result does not involve an expansion in the ratio $ \tau = m^2 /M^2 $.  
For such an expansion the following representation of the dilogarithm is useful
\begin{align}
{\rm Li}_2 (1 - \tau ) = \frac{\pi^2}{6}  + \sum_{n=1}^\infty  \left[ \frac{\ln \tau}{n} - \frac{1}{n^2}  \right] \tau^n \quad (0 < \tau < 1) 
\fs
\end{align}
In the case of equal masses, $m = M $, the result reduces to the simpler expression \cite{Broadhurst:1991fi, Gasser:1998qt}
\begin{align}
\bar{S}_{1,1,1} \{  \M,\M, \M ; \M^2\} = -  \frac{\M^2} {(4\pi)^4}   \left[   \frac{\pi^2}{4} + \frac{15}{8} \right]
\fs 
\end{align}
For the case of two equal masses and zero momentum, $s= 0 $, one finds \cite{vanderBij:1983bw,Davydychev:1992mt,Caffo:1998du}
\begin{align}
\bar{S}_{1,1,1} \{M ,M, m ; 0 \} & = \frac{M^2}{(4\pi)^4}  \left\{  \frac{4- \tau}{2 } \, F(\tau)
+\frac{\tau}{2} \ln^2 \tau  - (2 +\tau )\left[   \frac{\pi^2}{12} + \frac{3}{2} \right]  \right\} 
\co 
\end{align}
as above, $\tau = m^2/M^2$. The function $F(\tau)$ is given by
\begin{align}
\label{F(tau)}
F(\tau)  & =  \frac{1}{\sigma} \left[ 4 {\rm Li}_2 \left( \frac{\sigma-1}{\sigma+ 1} \right) + \ln^2 \left( \frac{1-\sigma}{1+\sigma} \right) + \frac{\pi^2}{3} \right] 
\co \quad 
\sigma = \sqrt{1-\frac{4}{\tau} } 
\fs 
\end{align}
$F(\tau)$ is real-valued on the 
positive real axis. In the range $ 0 < \tau < 4 $ where the variable $\sigma $ is purely imaginary it is given by 
\begin{align}
F(\tau)  & = 4  \sqrt{\frac{\tau}{4 -\tau}} \,{\rm Im}\, {\rm Li}_2 \left(e^{i 2 \arctan \sqrt{\frac{\tau}{4 -\tau}}} \right)  = 4  \sqrt{\frac{\tau}{4 -\tau}} \, {\rm Cl}_2 \left( 2 \arctan \sqrt{\frac{\tau}{4 -\tau}} \right) 
\fs   
\end{align}      
where we have introduced the Clausen function 
${\rm Cl}_2(x) $ \cite{Lewin}. 

\section{The pion mass to two loops in 3-flavour chiral perturbation theory}
\label{pion mass}

At leading order in the chiral expansion, the masses of $ \pi $, $ K $ and $ \eta $ are given by \cite{Gasser:1984gg}
\begin{align}
\mpi^2 = 2 \mh B_0 
\co \quad 
\mK^2 = (m_s+ \mh ) B_0  
\co \quad 
\meta^2  =  \frac{2}{3} (2 m_s +\mh) B_0 
\fs 
\end{align}
As they stand, the formulas are valid in the isospin limit $ m_u =m_d =\mh$. The three masses satisfy the Gell-Mann-Okubo relation
\begin{align}
\meta^2    =  \frac{1}{3} ( 4  \mK^2 -   \mpi^2 )  \fs 
\end{align} 
The chiral expansion of the physical pion mass $M_\pi^2$ can be written as
\begin{align}
M_\pi^2 = m_\pi^2 \{ 1  + \delta_\pi^{(2)} + \delta^{(4)}_\pi  \} + O(p^8) 
\fs
\end{align}
with $\delta_\pi^{(n)}= O(p^{n})$. The corrections of relative order $p^2$ were determined in Ref.~\cite{Gasser:1984gg}, 
\begin{align}
\delta^{(2)}_\pi & = \frac{1}{F_0^2} \left[  \, 
m_\pi^2 \Lpi - \tfrac{1}{3}  m_\eta^2 \Leta  - 12 ( \mpi^2 + \meta^2 )(L^r_4 - 2 L^r_6)  -  8 \mpi^2 (L^r_5 - 2 L^r_{8} )  \, 
 \right] 
\co
\end{align} 
where the symbols $\Lpi $ and $\Leta $ denote the chiral logarithms
\begin{align}
\LP=\frac{1}{2(4\pi)^2} \ln \frac{\mP^2}{\mu^2} \co \quad P =\pi, \, K, \, \eta 
\fs
\end{align} 

At the two-loop level, using the representation for the pion mass given in 
Ref.~\cite{Amoros:1999dp},\footnote{We have checked that the representation for the pion mass as printed in Ref.~\cite{Amoros:1999dp} coincides with the representation given in the preprint version of Ref.~\cite{Golowich:1997zs}, provided the tree contributions from the order $p^6$ Lagrangian are identified properly.
} 
we find the following contributions
\begin{align}
\label{delta decomposition}
\delta^{(4)}_\pi = \frac{1}{\Fz^4}  \left[  \,  \cloop + \cloglog + \clog + \clogLi + \cLi + \cLiLj + \cCi\,   \right] 
\fs 
\end{align}
The  term $\cCi$ is given by the tree contributions of the order $p^6$ Lagrangian \cite{Bijnens:1999sh}\footnote{
Our normalization conventions for the $C^r_i$ are such that they have mass dimension $-2$ and the canonical large $N_c$ behaviour.}
\begin{align}
 & ( 64 \Fz^2)^{-1} \cCi =  -\mK^4 [C^r_{16} - C^r_{20} - 3 C^r_{21} ] - \mK^2 \mpi^2 [ C^r_{13}  + \tfrac{1}{2} C^r_{15} - C^r_{16} - 3 C^r_{21} - C^r_{32} ]
\\
& - \tfrac{1}{4} \mpi^4 [2 (C^r_{12}+ C^r_{13})  + C^r_{14}  +  C^r_{15} + 3  C^r_{16}  +C^r_{17}  -3  C^r_{19} - 5 C^r_{20} -  3 C^r_{21} -2 ( C^r_{31}+ C^r_{32}) ]
\co \non
\end{align}
while the contributions bilinear in the $L_i$ are collected in the term $\cLiLj$,
\begin{align}
\label{cLixLj}
\cLiLj & = 64 \, [ (2 \mK^2 + \mpi^2) L^r_4 + \mpi^2 L^r_5] \,  [ (2 \mK^2 + \mpi^2) (L^r_4 -2 L^r_6) + \mpi^2 (L^r_5-2 L^r_8)]
\fs
\end{align}
Further, the quantity $\cLi $ accounts for the terms linear in the $L_i$ and involving no chiral  logarithms, 
 \begin{align}
9(4\pi)^2 \cLi  & = 
\mK^4  \left[  104 L^r_2  + \tfrac{86}{3} L_3 + 32 L^r_4 + \tfrac{64}{3} L^r_5 -64 (L^r_6 + L_7 +L^r_8)  \right] 
\\ 
& \quadp - \mK^2 \mpi^2  \left[  16 L^r_2  + \tfrac{16}{3} L_3 + 64 L^r_4 + \tfrac{32}{3} L^r_5 -64 (2L^r_6 + 2L_7 +L^r_8)  \right] 
\no
& \quadp  + \mpi^4  \left[  36 L^r_1 + 74 L^r_2  + \tfrac{56}{3} L_3 - 40 L^r_4 - \tfrac{104}{3} L^r_5 + 80 L^r_6 -64  L_7 + 48 L^r_8  \right] 
 \co\non 
\end{align}
while the products chiral logarithm times $ L_i$ are collected in the term $ \clogLi $, 
\begin{align}
\clogLi  & =  
\mK^4  [  - 32  ( 4L^r_1 + L^r_2  + \tfrac{5}{4} L_3 - 4  L^r_4 -  L^r_5 + 4 L^r_6 + 2 L^r_8 ) \, \LK \quad 
\\
 & \quadp  - \tfrac{128}{9} ( 4L^r_1 + L^r_2  +  L_3 -\tfrac{11}{2} L^r_4 -  L^r_5 + 6  L^r_6 - 3 L_7 +  L^r_8 ) \, \Leta  ] 
\no
& \quadp - \mK^2 \mpi^2  [  16  ( 3 L^r_4 - 4 L^r_6)   \Lpi 
-  \tfrac{64}{9} ( 4L^r_1 + L^r_2  +  L_3  -\tfrac{13}{4} L^r_4  + 3  L^r_6 - 6  L_7 -2  L^r_8 ) \, \Leta  ] 
\no
 & \quadp + \mpi^4 [  - 8  ( 14 L^r_1 + 8 L^r_2  + 7  L_3 - 9   L^r_4 - 3  L^r_5 + 16 L^r_6 + 6 L^r_8 ) \, \Lpi   
\no
 & \quadp  - \tfrac{8}{9} ( 4L^r_1 + L^r_2  +  L_3 -  L^r_4 +  L^r_5 -2   L^r_8 ) \, \Leta  ] 
\fs \non 
\end{align} 
The term $\clog$ collects the contributions which involve single chiral logarithms
\begin{align}
(4\pi)^2 \clog & = - \frac{76  \mK^4  }{81}[ 9 \LK + \Leta]  - \frac{\mK^2 \mpi^2}{36}    [ 47 \Lpi + 12 \LK + \tfrac{85}{9} \Leta ] - \frac{\mpi^4}{48} [ 373 \Lpi - \tfrac{161}{27} \Leta ]
\co \non 
\end{align}
and $\cloglog$ is given by the sum of the terms bilinear in the chiral logarithms, 
 \begin{align}
 \cloglog & = \mK^4 \left[  \tfrac{175}{36} (\LK)^2 - \tfrac{59}{18} \LK \Leta + \tfrac{25}{12} (\Leta)^2 \right] +  \frac{\mpi^4}{9} \left[  \tfrac{337}{4} (\Lpi)^2 + 5  \Lpi \Leta - \tfrac{1}{12}  (\Leta)^2 \right] 
 \no 
 & \quad +  \mK^2  \mpi^2 \left[  \tfrac{3}{2} (\Lpi)^2  -  \Lpi ( \LK + \tfrac{8}{9} \Leta) + \tfrac{1}{9} ( 2 (\LK)^2 + 5 \LK \Leta  - \tfrac{25}{6} (\Leta)^2 ) \right]
 \fs  
\end{align}
Making use of the known renormalization scale dependence of the low-energy constants $L^r_i$ and $C^r_i$ \cite{Gasser:1984gg,Bijnens:1999hw}, one readily checks that the sum of the terms given so far is scale independent.   
Finally, the term $\cloop$ accounts for 
the contributions from the sunset integrals in a scale independent manner. It can further be decomposed as follows,
\begin{align}
\cloop & = - \frac{1}{(4\pi)^4} 
\left\{    \mK^4 \left[  \frac{97 \pi^2}{432} - \frac{23}{96}  \right]  
+\mK^2 \mpi^2  \left[   \frac{25 \pi^2}{324}+\frac{35}{432}  \right]  
+ \mpi^4 \left[  \frac{41 \pi^2}{1296} -\frac{4429}{3456}   \right]  
\right\} 
\no 
& \quadp + \bar{c}_{KK\pi}+  \bar{c}_{\eta\eta\pi} +  \bar{c}_{KK\eta} 
\fs
\end{align}
The indices refer to the particle content in the contributing sunset graphs. Explicitly, we have
\begin{align}
\bar{c}_{KK\pi}  & = 
\left[ \frac{\mK^4}{8 \mpi^2} - \frac{3 \mK^2}{4}  - \frac{3 \mpi^2}{8} \right] \, \bar{S}_{K,K,\pi} ( \mpi^2 ) 
+ \frac{\mK^2(\mK^4 - \mpi^4)}{2 \mpi^2}  \, \bar{S}_{2 K,K,\pi} ( \mpi^2 )  
\\
& \quadp  - \frac{1}{(4\pi)^4} \left\{  \frac{\mK^6}{\mpi^2} \left[  \frac{\pi^2}{48}  + \frac{3}{8}  \right] + \frac{\mK^4}{16} \, \ln \frac{\mpi^2}{\mK^2}  \left[ \ln \frac{\mpi^2}{\mK^2}  + 1 \right]   \right\}
\co \no
\bar{c}_{\eta\eta\pi} & =  -\frac{\mpi^2}{18} \, \bar{S}_{\eta,\eta,\pi} ( \mpi^2) 
\co \no
\bar{c}_{KK\eta} & = 
\left[ \frac{5 \mK^4}{8 \mpi^2} - \frac{43 \mK^2}{36}  + \frac{17 \mpi^2}{72} \right] \, \bar{S}_{K,K,\eta} ( \mpi^2 ) 
+ \left[ \frac{4 \mK^6}{3 \mpi^2} - \frac{5 \mK^4}{3}  + \frac{1}{3} \mK^2 \mpi^2  \right] \,   \bar{S}_{2 K,K,\eta} ( \mpi^2 )  
\no 
&  \quadp +  \left[ \frac{2 \mK^6}{3 \mpi^2} - \frac{65 \mK^4}{54}  + \frac{17}{27} \mK^2 \mpi^2  - \frac{5 \mpi^4}{54}  \right] \,   \bar{S}_{K,K,2 \eta} ( \mpi^2 )  
\no
& \quadp  + \frac{1}{(4\pi)^4} \left\{  \frac{\mK^6}{\mpi^2} \left[   \frac{\pi^2}{144} +\frac{1}{8}  - \frac{1}{12} \, \ln^2  \frac{\meta^2}{\mK^2}  \right] 
-  \frac{5 \mK^4}{16} \, \ln \frac{\mpi^2}{\mK^2}  \right\}
\fs \non
\end{align}
In the above equations, we have introduced an abbreviated notation for the sunset master integrals 
\begin{align}
 \bar{S}_{a P,b  Q ,c  R } (s)  =  \bar{S}_{a,b,c} \{ m_P, m_Q ,m_R; s \}  
 \fs
\end{align} 
Note that the splitting into the various contributions is not 
unique and neither is the choice of the scale independent piece $\cloop$. Our choice has the property that all the pieces do separately have a regular behaviour in the limit $ \mpi^2 \to 0 $. For $\bar{c}_{KK\pi}$ and $\bar{c}_{\eta\eta\pi}$ this follows from the explicit result for the sunset integral given in Eq.~(\ref{Eq:Sunset MMm on-shell}). For the contribution $\bar{c}_{KK\eta}$ we do not have a closed expression. However, the results from Appendix~\ref{App:Sunset for s=0} imply that the expansion of this term around $\mpi^2 = 0$ is determined by the function $\bar{S}_{1,1,1} \{ 1,1, \tfrac{2}{\sqrt{3}}; 0 \}$ evaluated for zero pion mass. Explicitly, we find
\begin{align}
\label{cKKetaExp}
& \frac{(4\pi)^4}{\mK^4}\,  \bar{c}_{KK\eta} = 
-\frac{2}{3} \, F(\tfrac{4}{3}) - \frac{43}{144} \, \ln^2 \frac{4}{3} + \frac{95 \pi^2}{864} +\frac{425}{192} 
\\
& \quad + \left[  -\frac{43}{72} \, \ln \rho + \frac{113}{288} \, F(\tfrac{4}{3}) +\frac{1}{24} \, \ln^2 \frac{4}{3} +\frac{35}{288} \, \ln \frac{4}{3}  - \frac{ \pi^2}{54} -\frac{15}{32} \right]   \rho
\no
& \quad  + \left[  \frac{17}{144} \, \ln \rho - \frac{5725}{147456} \, F(\tfrac{4}{3}) +  \frac{1}{144} \, \ln^2 \frac{4}{3} - \frac{13}{4096} \, \ln \frac{4}{3}  - \frac{ \pi^2}{864} 
- \frac{1217}{9216} \right]   \rho^2 +O(\rho^3)
\co \non
\end{align} 
where $ \rho = \mpi^2 / \mK^2 $ and the function $F(x)$ is defined in Eq.~(\ref{F(tau)}). On the basis of the numerical values given in Table~\ref{num MI} one can verify that the deviation between the full result and the approximation is inferior to $10^{-4}$ for the physical value of $ \rho$. In fact, even for $ \rho = 1 $ the three terms given explicitly still account for more than $99\% $ of the full result. 
We conclude that in the whole range of interest Eq.~(\ref{cKKetaExp}) is also valid for studies of the quark mass dependence of $\bar{c}_{KK\eta}$ as needed e.g. for calculations in lattice QCD.\footnote{We thank L.~Lellouch for helpful discussions on the subject.}

%
%
\begin{table}[t]
\renewcommand{\arraystretch}{1.4}
\begin{center}
\begin{tabular}{crrrr}
\hline
						& $\{1,1,1;1\}$ 		&$\{1,1,\sqrt{\rho};\rho\}$		&$\{\sqrt{\lambda},\sqrt{\lambda},\sqrt{\rho};\rho\}$ 	&$\{1,1,\sqrt{\lambda};\rho\}$		
\\ \hline
 $(4\pi)^4 \bar{S}_{1,1,1}$		& $-4.34240$		&	$-4.03404$			& $-5.38100$								&	$-4.03912$			
\\
$(4\pi)^4 \bar{S}_{2,1,1}$		& $1.32247$		&	$2.16682$			& $2.19634$ 								&	$0.935212$			
\\
$(4\pi)^4 \bar{S}_{1,1,2}$		& 			         &	$-5.49841$			& $ -6.53194$								&	$1.55782$			
\\ \hline
\end{tabular}
\caption{
\label{num MI}
Approximate numerical values for various master type integrals, indices and arguments as indicated. 
The quantity $ \sqrt{\rho}$ denotes the ratio of the physical pion and kaon masses, $ \rho = M_\pi^2/M_K^2 \simeq 0.0743454$.
$ \sqrt{\lambda}$ stands for the corresponding ratio of the $\eta$ and kaon masses, in the approximation where the $\eta$ mass 
is expressed through the Gell-Mann-Okubo relation, $\lambda= \tfrac{1}{3}(4  - \rho ) \simeq 1.30855$. 
}
\end{center}
\renewcommand{\arraystretch}{1.0}
\end{table}
%
%

As a check of our result for the two-loop correction given in the present section, we have verified that its expansion for $\mh/m_s \ll1 $ matches the two-loop representation for the pion mass in two-flavour chiral perturbation theory \cite{Burgi:1996qi,Bijnens:1997vq,Bijnens:1998fm}. Furthermore, we have thereby recovered the matching relations for the two-flavour low-energy coupling constants $ B$ and $ \ell^r_3 $ \cite{Gasser:1983yg} given in Refs.~\cite{Kaiser:2006uv} and \cite{Gasser:2007sg}, respectively. As a final remark we note that in the SU(3) limit, $ \mpi = \mK = \meta = \mP$, the contributions from the sunsets simplify quite dramatically to yield
\begin{align}
\cloop  = \frac{1457}{384}  \frac{\mP^4}{(4\pi)^4}
\co \quad 
\cloglog = \frac{27 \mP^4 }{2} \,(\LP)^2 
\co \quad 
\clog=  -\frac{1363}{72} \frac{\mP^4}{(4 \pi)^2} \,  \LP
\fs 
\end{align}  
 
\section{Numerical analysis}
\label{numerics}

We conclude the present paper with a discussion of the numerical implications of our results.
For an investigation of the size of the various contributions it is advantageous to re-write the corrections in terms of the physical masses $\Mpi$, $ \MK$, 
$\Meta$
and the physical pion decay constant $F_\pi$ \cite{Gasser:1984gg}.
We denote the corresponding corrections by the capital letter $\Delta$ and write
\begin{align}
\label{Def: capital Deltas}
M_\pi^2 = m_\pi^2 \{ 1  + \Delta_\pi^{(2)} + \Delta^{(4)}_\pi  \} + O(p^8) 
\fs
\end{align}
The requirement $\Delta_\pi^{(2)} -  \delta_\pi^{(2)} = O(p^4) $ does not determine the form of $\Delta_\pi^{(2)} $ uniquely. We make the choice
\begin{align}
\label{Delta2pi}
\Delta_\pi ^{(2)} & = \frac{1}{\Fpi^2} \left[  
\, \Mpi^2 \LLpi - \tfrac{1}{3}  \Meta^2 \LLeta 
 - 12 ( \Mpi^2 + \Meta^2 )(L^r_4- 2 L^r_6 )  - 8 \Mpi^2 (L^r_5- 2 L^r_{8}) \, 
 \right] 
\co
\end{align} 
where $\LLpi $ and $\LLeta $ denote the chiral logarithms involving the physical meson masses
\begin{align}
\LLP =\frac{1}{2(4\pi)^2} \ln \frac{\MP^2}{\mu^2} \co \quad P =\pi, \, K, \, \eta 
\fs
\end{align} 
The expression in Eq.~(\ref{Delta2pi}) is thus is strictly independent of the renormalization scale $\mu$,
\begin{align}
\mu \frac{d}{d\mu} \Delta_\pi ^{(2)} = 0 
\fs
\end{align}
Note that in the formula above the contribution from the combination $L^r_4- 2 L^r_6$ is enhanced by a factor of about 26 relative to the one from $L^r_5- 2 L^r_8$.

The next-to-next-to leading order correction $\Delta^{(4)}_\pi$ is given by the contributions from $\delta^{(4)}_\pi$ plus a shift generated by the difference $\delta^{(2)}_\pi -\Delta^{(2)}_\pi $.
It allows a decomposition analogous to the one in Eq.~(\ref{delta decomposition}).
The corresponding numerical\footnote{In the numerical evaluation, we use the Dashen-corrected values of the physical meson masses, 
$ \Mpi = M_{\pi^0} \simeq 134.98\,  \mbox{MeV} $, 
$  \MK = \sqrt{ \frac{1}{2}(M_{K^+}^2 + M_{K^0}^2 - M_{\pi^+}^2 + M_{\pi^0}^2) } \simeq 495.03 \,  \mbox{MeV} $, 
$ M_\eta = 547.51\,  \mbox{MeV} $ and $ \Fpi = 92.42\, \mbox{MeV}$  \cite{Yao:2006px}.}  
 contributions are listed in Table~\ref{table:Delta_pi_4}. To show the variation with the scale we give all values for three different choices of the renormalization scale $\mu$. For the terms involving the order $p^4$ couplings constants 
$L_1^r, \ldots,  L_8^r$,
we used the numerical values provided in Ref.~\cite{Gasser:1984gg} and fit 10 of Ref.~\cite{Amoros:2001cp}, respectively. While the 
two sets of values have been determined on the basis of an order $p^4$ \cite{Gasser:1984gg} and an order $p^6$ \cite{Amoros:2001cp} phenomenological analysis, respectively,
the difference between the resulting numbers is rather small. The following observations can be made: The tree graph contributions bilinear in the $L_i$ are small throughout and amount to less than
$1.2 \cdot 10^{-3}$ for both sets of the $L_i$ and all three scales $ \mu $. The contribution linear in the $L_i$ and involving no logs is given by
\begin{align}
\hspace*{-0.7em}\frac{2}{9 (4\pi)^2  \Fpi^4  } \left[ 
\MK^4  \left( 52 L^r_2  + \tfrac{43}{3} L_3 \right)
- \MK^2 \Mpi^2   \left( 8  L^r_2  + \tfrac{8}{3} L_3 \right) 
+ \Mpi^4  \left(  18 L^r_1 + 37 L^r_2  + \tfrac{28}{3} L_3  \right)  \right] .
\end{align}
For this combination the resulting values of Refs.~\cite{Gasser:1984gg} and \cite{Amoros:2001cp} happen to be close despite the fact that the values for the individual coupling constants differ considerably.
At the scale $ \mu = 770 \, \mbox{MeV}$ the bulk of the contribution to $\Delta^{(4)}_\pi$ comes from two-loop contributions not involving the $L_i$. Among those terms, the one from the single logs is dominant, and the scale independent contributions from the sunset graphs are also sizeable. On the other hand, the contribution from the double chiral logs is small -- those terms alone only yield a poor approximation to the full result. If the two contributions of type $ \log \times \log $ and $ \log \times L_i $ are combined this leads to partial cancellations in both, the size of the contributions and their scale dependence. 

\begin{table}[t]
\renewcommand{\arraystretch}{1.4}
\begin{center}
\begin{tabular}{ccccccccc}
\hline
$\mu $      			& $L_i^r$                          													
& $ \overline{{\rm loop}} $ &  $ \log \times \log $  & $ \log$ & $\log  \times L_i  $ 	&  $L_i $								&$ L_i \times L_j $  							& $ \Sigma  $ 	
\\ \hline  
$ M_\eta$ 		&  \begin{tabular}{l}Ref.~\cite{Gasser:1984gg}\\Ref.~\cite{Amoros:2001cp}\end{tabular}  
& $7.60$	& $0.94$ 		& $3.33$ 	& \begin{tabular}{c}$-0.56$\\$0.61$\end{tabular}	&  \begin{tabular}{c}$2.94$\\$2.93$\end{tabular}	& \begin{tabular}{c}$0$ \\ $ 0.11$ \end{tabular}	& \begin{tabular}{c}$14.3$ \\ $15.5$ \end{tabular}	
\\ \hline                              
$ 770\, {\rm MeV}  $ 	&  \begin{tabular}{l}Ref.~\cite{Gasser:1984gg}\\Ref.~\cite{Amoros:2001cp}\end{tabular}   
& $7.60$	& $3.89$		& $14.0$ 	& \begin{tabular}{c}$-3.91$\\$-0.29$\end{tabular}	&  \begin{tabular}{c}$0.52$\\$0.51$\end{tabular}	& \begin{tabular}{c}$-0.01$\\$-0.002$\end{tabular}		& \begin{tabular}{c}$22.1$ \\ $25.7$\end{tabular}		
\\ \hline              
$ 1 \,{\rm GeV}$ 	& \begin{tabular}{l}Ref.~\cite{Gasser:1984gg}\\Ref.~\cite{Amoros:2001cp}\end{tabular}   
& $7.60$	& $8.60$ 		& $22.2$ 	& \begin{tabular}{c}$-11.4$\\$-5.88$\end{tabular}	&  \begin{tabular}{c}$-1.34$\\$-1.35$\end{tabular}	& \begin{tabular}{c}$0.07$\\$0.001$\end{tabular}		& \begin{tabular}{c}$25.7$ \\ $31.2$\end{tabular}		
\\ \hline
\end{tabular}
\caption{
\label{table:Delta_pi_4}
Numerical contributions
 (in units of $10^{-2}$) to the 2-loop correction $\Delta^{(4)}_\pi$ as defined in Eq.~(\protect\ref{Def: capital Deltas}), evaluated for three different choices of the renormalization scale $ \mu $.
Note that the tree contributions from the order $p^6$ Lagrangian are absent. The other contributions are listed separately. 
For the $L_i$ we use the numerical central values taken from Ref.~\cite{Gasser:1984gg} and fit 10 of Ref.~\cite{Amoros:2001cp}, respectively, as indicated. 
The last column gives the sum of the contributions in the other columns. 
We repeat that this sum does not account for the contributions of the low energy constants $C_i$ (see text).}
\end{center}
\renewcommand{\arraystretch}{1.0}
\end{table}

We add a remark concerning the contributions from the coupling constants $C_i$ of the order $p^6$ chiral Lagrangian. The contribution to $ \Delta_\pi^{(4)}$ from the $ C_i $ consists of three terms accompanied by the factors $ \MK^4 $, $ \MK^2 \Mpi^2 $ and $ \Mpi^4 $, respectively. The results of Ref.~\cite{Cirigliano:2006hb} imply a non-vanishing contribution only to the third term. In this respect, this contribution is similar to the tree result in Eq.~(\ref{cLixLj}) where the only contribution surviving the large $ N_c$ limit is proportional to $  L_5^r (L_5^r - 2 L_8^r) /\Fz^4 $ whereas the rest of the terms all involve a $1/N_c$ suppressed factor of $L_4^r /\Fz^2$ or $ L_6^r /\Fz^2  $. The non-vanishing estimate for the relevant combination of $C_i$ from 
Ref.~\cite{Cirigliano:2006hb} involves unknown contributions related to resonance matrix elements of scalar and pseudoscalar resonances. On general grounds~\cite{Cirigliano:2005xn} one expects that the resulting contributions are of the same nature as e.g. the one for $(L^r_5)^2$.
In view of the small ratio $ \Mpi/\M_S$, this contribution amounts numerically to very little,
\begin{align}
\frac{4 \Mpi^4}{M_S^4} \simeq 10^{-3}  \quad \mbox{for}  \quad M_S = 1 \,\mbox{GeV} 
\fs 
\end{align}
On the basis of these observations one would not expect a sizeable contribution to $ \Delta_\pi^{(4)}$ from the coupling constants $C_i$. This is to be confronted with the substantial scale dependence of the terms given in Table~\ref{table:Delta_pi_4}. Assuming that the contributions of the $C_i$ vanish somewhere in the range $ \mu = \Meta \ldots 1\,\mbox{GeV} $ leads to an estimate of $ \Delta_\pi^{(4)} $ between $0.14$ and $0.31$, depending also on the preferred set of the $L_i$. 

\section{Conclusions}
\label{Conclusions}

In the present paper we reviewed the recurrence relations for the sunset integral in the general mass case. In Section~\ref{tensorial sunsets}, we in particular provide the explicit representation for the tensorial integrals of rank 1 and 2 in terms of the master integrals. In Section~\ref{Renormalization and results}, we discuss the renormalization of the sunset master integrals and provide a collection of known results for the finite parts of the master integrals.    

In Section~\ref{pion mass}, we used these results to obtain a simplified representation of the pion mass to two loops in three-flavour chiral perturbation theory on the basis of the result provided in Ref.~\cite{Amoros:1999dp}. Our final result involves 6 master integrals accounting for the contributions of intermediate $KK\pi$, $\eta\eta\pi$ and $ KK\eta$ states while the contributions from the intermediate 3-pion states have been evaluated and are not displayed explicitly. The result of Ref.~\cite{Amoros:1999dp} was given in terms of 10 functions corresponding to scalar and tensorial sunset integrals. 
Besides the smaller total number of basis functions, our choice has the advantage that a nontrivial calculation is only once needed for each intermediate state. Once the unit index integrals are known explicitly, the remaining master integrals can be obtained by derivatives with respect to the masses. Of those three master integrals the two for the $KK\pi$ and $ \eta \eta \pi $ intermediate states are determined by the same known function given in Section~\ref{Renormalization and results}. For the master integral involving a $KK\eta$ intermediate state we do not have an explicit expression. However, its expansion for $\Mpi^2 \to 0 $ can be given in terms of the function evaluated for zero pion mass (which is also provided in Section~\ref{Renormalization and results}) and we obtain a numerically very accurate approximation of the function. We assume that this representation will be suitable for all practical purposes including investigations of the quark mass dependence of the two loop contribution to the pion mass. To know the closed-form result would nevertheless be interesting. Given our representation this will be the case once the analytic result for the sunset integral with two equal masses is known.

In chiral perturbation theory, the genuine two-loop contributions only represent part of the final result at order $p^6$. In addition, there arise one loop graphs with insertions of the order $p^4$ coupling constants $L_i$. Further, there are tree contributions bilinear in the $L_i$ or involving contributions from the coupling constants $C_i$ of the order $p^6$ chiral Lagrangian. Unfortunately, the numerical values for the latter are presently not known. 
In Section~\ref{numerics}, we show that a simple estimate on the basis of resonance saturation indicates very small contributions from these coupling constants. 
Neglecting them altogether yields an estimate of a total order $p^6$ contribution $ \Delta_\pi^{(4)} $ between $0.14$ and $0.31$, where the main uncertainty is due to the choice of the 
renormalization 
scale at which the contributions from the $C_i$ are supposed to vanish. Qualitatively, this result for a rather sizeable correction at two-loop order is in agreement with earlier findings in Refs.~\cite{Amoros:1999dp,Amoros:2001cp,Amoros:1999qq,Amoros:2000mc,Bijnens:2002hp,Bijnens:2003uy,Bijnens:2003xg,Bijnens:2004eu,Bijnens:2004bu,Bijnens:2007si}.

Further studies will be required to arrive at a conclusive understanding of the nature of the three-flavour chiral expansion. We are convinced that our explicit representation for the pion mass at two loop order will be helpful in forthcoming investigations. For instance, knowledge of the quark mass dependence of the two-loop corrections will ultimately be needed for studies in lattice QCD. Naturally, the methods and results discussed in the present paper will also be applicable to other observables in three flavour chiral perturbation theory.
For a future more comprehensive phenomenological study it would of course be desirable to have similarly simplified representations for further of these.

\section*{Acknowledgements}

We are grateful to 
T.~Becher,  
G.~Ecker, 
J.~Gasser,
D.~Greynat, 
C.~Haefeli, 
H.~Leutwyler,
E.~de Rafael, 
J.~Schweizer and 
P.~Talavera for suggestions and comments on the manuscript. 
This work was supported by the Swiss National Science Foundation.

\begin{appendix}

\section{Recurrence relations for the one-loop 2-point function}
\label{App:recrel-one-loop}

For the purpose of illustration we list  in this appendix the recurrence relations for the one-loop 2-point function defined by
\begin{align}
\label{I_ab Def}
I^{d}_{\alpha,\beta}  \{ \M_1 , \M_2 ; p^2  \} & =  \frac{1}{i} \!  \int \! \frac{d^d k}{(2\pi)^d} \frac{1}{[M_1^2-k^2-i \eps]^\alpha} \frac{1}{[M_2^2-(p-k)^2-i \eps]^\beta} 
\fs
\end{align}
The corresponding Feynman parameter representation is
\begin{align}
\label{I_ab FP}
& I^{d}_{\alpha,\beta}  \{ \M_1 , \M_2 ; s \} = 
\\
& \quad \frac{1}{(4\pi)^\frac{d}{2}} \frac{\Gamma(\alpha+\beta-\frac{d}{2})}{\Gamma(\alpha)\Gamma(\beta)} 
\int_0^1 \! dx \, x^{\alpha -1} \bar{x}^{\beta -1}   [ x  \M_1^2 + \bar{x}  \M_2^2 - x \bar{x}\, s- i \eps ]^{\frac{d}{2}-\alpha-\beta} 
\co \non
\end{align}
with $\bar{x} =1-x$. In the case of the one-loop integral the recurrence relations follow immediately from the integration by parts relations associated with the two available momenta $p$ and $k$. Evaluating these for the case of the integral $ I^d_{\alpha,\beta} \{ \M_1 , \M_2 ; s  \} $ in Eq.~(\ref{I_ab Def}) leads to two equations involving in particular $I^d_{\alpha+1 ,\beta}  \{ \cdot  \}  $ and $I^d_{\alpha ,\beta+1}  \{ \cdot  \} $. Solving for the former yields  
\begin{align}
& I^{d}_{\alpha+1,\beta}  \{ \M_1 , \M_2 ; s  \}  = 
\frac{1}{\lambda(\M_1^2,\M_2^2,s)} \bigg\{  
(\M_1^2 + \M_2^2-s) I^{d}_{\alpha+1,\beta-1}  \{ \cdot  \} 
\\
& \quad \quad \quad -  \frac{(d-\alpha-2\beta)(\M_1^2-s)-(d-3\alpha)\M_2^2}{\alpha} \,  I^{d}_{\alpha,\beta}  \{ \cdot  \} 
- \frac{2 \beta \M_2^2}{\alpha} \, I^{d}_{\alpha-1,\beta+1}  \{ \cdot  \}  
\bigg\} 
\co \non 
\end{align}
with $\lambda (x,y,z)  = x^2 + y^2 +z^2 -2 x y - 2y z -2z x$.  The second relation can be obtained by the interchange $(\alpha,\M_1) \leftrightarrow (\beta , \M_2)$. The r.h.s of the above equation only involves integrals $ I^d_{a,b} \{ \cdot \} $ with $a+b = \alpha + \beta $. Therefore, all the integrals with positive integer indices can be reduced to the integral with unit indices and the tadpoles $ I^d_1\{ \M_1\} $ and $ I^d_1\{ \M_2\} $.   

Next, consider the tensorial Feynman integrals $ I^{(n)d}_{\alpha,\beta} $ defined by 
\begin{align}
\label{I_ab Tensors}
 \frac{1}{i} \!  \int \! \frac{d^d k}{(2\pi)^d} \frac{\tau^{\mu_1 \cdots  \mu_n}(k)}{[M_1^2-k^2-i \eps]^\alpha} \frac{1}{[M_2^2-(p-k)^2-i \eps]^\beta}  =
\tau^{\mu_1 \cdots  \mu_n}(p)\,   I^{(n)d}_{\alpha,\beta}  \{ \M_1 , \M_2 ; p^2  \} \fs \!\!\!
\end{align}  
Evaluation via Feynman parameters leads to 
\begin{align}
I^{(n)d}_{\alpha,\beta}  \{ \M_1 , \M_2 ; s  \}  = (4\pi)^n \frac{\Gamma(\beta +n)}{\Gamma(\beta)} \, I^{d+2n}_{\alpha,\beta+n}  \{ \cdot \}
\fs   
\end{align}
A relation that allows to decrease the dimension of the integrals can be obtained from the Feynman parameter representation in Eq.~(\ref{I_ab FP}). Inserting $1 = x + (1-x)$ under the integral sign leads to 
\begin{align}
I^{d}_{\alpha,\beta}  \{ \M_1 , \M_2 ;s \} = 4\pi [ \, \alpha\, I^{d+2}_{\alpha+1,\beta}  \{\cdot  \} +  \beta \, I^{d+2}_{\alpha,\beta+1}  \{\cdot  \} \,]
\fs
\end{align} 
It suffices then to set $\alpha = \beta =1 $ and to apply the index recurrence relations to the r.h.s. to obtain the desired relation
\begin{align}
I^{d+2}_{1,1}   \{ \M_1 , \M_2 ; s \} = \frac{1}{d-1} \frac{1}{8 \pi s}  \bigg\{  
\lambda(\M_1^2, \M_2^2 , s) I^{d}_{1,1}  \{ \cdot  \}  & + (\M_1^2 - \M_2^2 +s )   I^{d}_1\{ \M_1  \} 
\\
& -  (\M_1^2 - \M_2^2 -s )   I^{d}_1\{ \M_2  \}  
\bigg\} 
\fs \non 
\end{align}
Applying the recurrence relations to the tensorial integrals of first and second rank, we reproduce the well-know results
\begin{align}
\label{I_ab Tensors reduced}
I^{(1)d}_{1,1}   \{ \M_1 , \M_2 ; s \} &  = \frac{\M_1^2- \M_2^2 +s}{2s} \, I^{d}_{1,1}   \{ \cdot \}  + \frac{1}{2s} 
[ I^{d}_{1}   \{ \M_1\}  - I^{d}_{1}   \{ \M_2\}  ]
 \co
\\
I^{(2)d}_{1,1}   \{ \M_1 , \M_2 ; s \} & =  \frac{d(\M_1^2- \M_2^2 + s)^2- 4 \M_1^2 s}{4(d-1)s^2} \, I^{d}_{1,1}   \{ \cdot \}  
+  \frac{d(\M_1^2 - \M_2^2 +s)}{4(d-1)s^2 }  I^{d}_{1} \{ \M_1\}  
\no
&  \quadp -  \frac{d(\M_1^2 - \M_2^2 )+ (3d-4) s}{4(d-1)s^2 }  I^{d}_{1}   \{ \M_2\} 
\fs \non 
\end{align}   
Of course, for the one-loop case where the tensorial integrals do not lead to non-reducible denominators, there is no need to apply the method described above. We repeat that the present appendix is primarily intended to serve as an illustration of the method which in the main text is applied to the sunset integrals. It is conceivable, however, that the method could be advantageous in an application involving tensorial integrals of high rank.

Independently of the method used to derive them, the reduced representations for the tensorial integrals (\ref{I_ab Tensors reduced}) appear to involve singularities in the limit $s \to 0$, while it is clear from their definition in Eq.~(\ref{I_ab Tensors}) that these must be absent. That this is indeed the case can be verified in the following manner: In the case of zero external momentum the master integral simplifies to
\begin{align}
I^{d}_{1,1}   \{ \M_1 , \M_2 ; 0 \}  = - \frac{I_1^d \{\M_1\}- I_1^d \{\M_2\}}{\M_1^2 -\M_2^2} \co 
\end{align}
which is sufficient to prove the absence of the leading singularities in Eqs.~(\ref{I_ab Tensors reduced}). The coefficients of the non-leading potentially singular terms involve also momentum derivatives of the master integral. However, these may again be re-expressed in terms of an integral with a shifted dimension
\begin{align}
 \frac{\partial^n }{\partial s^n} I^{d}_{\alpha,\beta}   \{ \M_1 , \M_2 ;s\}  =  (4\pi)^n \frac{\Gamma(\alpha+n)  \Gamma(\beta+n)}{\Gamma(\alpha)  \Gamma(\beta)} \, I^{d+2n}_{\alpha+n,\beta+n}   \{ \cdot \} 
 \fs   
\end{align}
For vanishing momentum, $s=0$, they are therefore all expressible in terms of tadpoles and one can establish the regular behaviour of any tensorial integral. 

\section{Recurrence relations for the sunset at zero momentum}
\label{App:Sunset for s=0}

If the external momentum vanishes, $s = 0$, there emerge simplified recurrence relations. The relation \cite{Tarasov:1997kx}
\begin{align}
S^{d}_{\alpha+1,\beta,\gamma} \{ \M_1, \M_2 ,\M_3 ; 0 \}  =  &\, \frac{1}{\alpha  \, \lambda( \M_1^2, \M_2^2 ,\M_3^2 )  }
\\  
 \times \bigg\{ 
& -  \left[ 2 (\alpha - \beta) \M_2^2  + (d-\alpha -2 \beta)( \M_1^2-   \M_2^2 - \M_3^2   ) \right] S^{d}_{\alpha,\beta,\gamma} \{ \cdot \} 
\no 
& -2 \beta   \M_2^2    \left[ S^{d}_{\alpha-1,\beta+1,\gamma} \{ \cdot \}  - S^{d}_{\alpha,\beta+1,\gamma-1} \{ \cdot \}  \right]  
\no 
 & +  \alpha ( \M_1^2+  \M_2^2 - \M_3^2   )  \left[ S^{d}_{\alpha+1,\beta-1,\gamma} \{ \cdot \}  - S^{d}_{\alpha+1,\beta,\gamma-1} \{ \cdot \}  \right]  \bigg\} 
\non
\end{align}
together with its permutations allow one to express any integral with positive integer indices in terms of the master integral with unit indices plus tadpoles. Likewise, there is also a simplified relation for shifting the dimension \cite{Tarasov:1997kx}
\begin{align}
S^{d+2}_{\alpha,\beta,\gamma} \{ \M_1, \M_2 ,\M_3 ; 0 \}  & = - \frac{1}{(4\pi)^2 d (d+2 -\alpha - \beta  -\gamma)}
\\
 \times  \bigg[  & \lambda( \M_1^2, \M_2^2 ,\M_3^2 ) \, S^{d}_{\alpha,\beta,\gamma} \{ \cdot \} 
+ (- \M_1^2+  \M_2^2 + \M_3^2   ) \, S^{d}_{\alpha-1,\beta,\gamma} \{ \cdot \}   
\no 
&    + ( \M_1^2-  \M_2^2 + \M_3^2   ) \, S^{d}_{\alpha,\beta-1,\gamma} \{ \cdot \}  
+ ( \M_1^2+  \M_2^2 - \M_3^2   ) \, S^{d}_{\alpha,\beta,\gamma-1} \{ \cdot \}  \bigg] 
\fs \non
\end{align}
Making use of these relations it is straightforward to show that the expressions for the tensorial sunset integrals in Eqs.~(\ref{S1}) and (\ref{S2}) do indeed have a regular expansion around $s=0$. Here, the relation for shifting the dimension comes into play because momentum derivatives of the integrals may be re-expressed as integrals with a shifted dimension: On the basis of the Feynman parameter representation one verifies the following general relation for the derivatives with respect to the masses or momentum
\begin{align}
 & \left(-\frac{\partial}{\partial \M_1^2}\right)^{m_1}  \left(-\frac{\partial}{\partial \M_2^2}\right)^{m_2}
   \left(-\frac{\partial}{\partial \M_3^2}\right)^{m_3}  \left(\frac{\partial}{\partial s}\right)^n S^{d}_{\alpha,\beta,\gamma} \{ \M_1, \M_2 ,\M_3 ; s \} = 
   \\
 & \quad
   (4 \pi)^{2 n} \frac{\Gamma(\alpha+ m_1 + n) \Gamma(\beta+ m_2 + n)\Gamma(\gamma+ m_3 + n)}{\Gamma(\alpha) \Gamma(\beta)\Gamma(\gamma)}\,  S^{d+2n}_{\alpha+ m_1 + n,\beta+ m_2 + n,\gamma+ m_3 + n} \{\cdot \}
 \fs \non   
\end{align} 
Making use of the above results we easily establish the expansion of the master integral $ S^d_{1,1,1} \{ \mK , \mK, \meta ;\mpi^2 \}  $ around $ \mpi^2 = 0$,
\begin{align}
S^d_{1,1,1} \{ \mK , \mK, \meta ;\mpi^2 \}  = \mK^{2d-6} \big\{  S^d_{1,1,1} \{1, 1, \tfrac{2}{\sqrt{3}} ;0  \}   (1 - \frac{5(d+3)(d -3)}{32 d} \, \rho  ) \\
- \frac{3 (d-2)}{64 d} I^d_1 \{ 1 \}   \left[ (5d-9)   I^d_1 \{ 1 \} + (d-3)  I^d_1 \{ \tfrac{2}{\sqrt{3}}  \}  \right]   \rho + O(\rho^2 )
\big\} 
\co \non 
\end{align}
where  $\rho = \mpi^2/\mK^2 $. 

\end{appendix}


\end{document}